\documentstyle[12pt]{article}
\def\hybrid{\topmargin -20pt	\oddsidemargin 0pt
	\headheight 0pt	\headsep 0pt
	\textwidth 6.25in	
	\textheight 9.5in	
	\marginparwidth .875in
	\parskip 5pt plus 1pt	\jot = 1.5ex}

\hybrid

\def\baselinestretch{1.2}

\catcode`\@=11

\def\marginnote#1{}
\newcount\hour
\newcount\minute
\newtoks\amorpm
\hour=\time\divide\hour by60
\minute=\time{\multiply\hour by60 \global\advance\minute by-\hour}
\edef\standardtime{{\ifnum\hour<12 \global\amorpm={am}%
	\else\global\amorpm={pm}\advance\hour by-12 \fi
	\ifnum\hour=0 \hour=12 \fi
	\number\hour:\ifnum\minute<10 0\fi\number\minute\the\amorpm}}
\edef\militarytime{\number\hour:\ifnum\minute<10 0\fi\number\minute}
\def\draftlabel#1{{\@bsphack\if@filesw {\let\thepage\relax
   \xdef\@gtempa{\write\@auxout{\string
      \newlabel{#1}{{\@currentlabel}{\thepage}}}}}\@gtempa
   \if@nobreak \ifvmode\nobreak\fi\fi\fi\@esphack}
	\gdef\@eqnlabel{#1}}
\def\@eqnlabel{}
\def\@vacuum{}
\def\draftmarginnote#1{\marginpar{\raggedright\scriptsize\tt#1}}

\def\draft{\oddsidemargin -.5truein
	\def\@oddfoot{\sl preliminary draft \hfil
	\rm\thepage\hfil\sl\today\quad\militarytime}
	\let\@evenfoot\@oddfoot	\overfullrule 3pt
	\let\label=\draftlabel
	\let\marginnote=\draftmarginnote
   \def\@eqnnum{(\theequation)\rlap{\kern\marginparsep\tt\@eqnlabel}%
\global\let\@eqnlabel\@vacuum}  }

\def\preprint{\twocolumn\sloppy\flushbottom\parindent 2em
	\leftmargini 2em\leftmarginv .5em\leftmarginvi .5em
	\oddsidemargin -.5in	\evensidemargin -.5in
	\columnsep .4in	\footheight 0pt
	\textwidth 10.in	\topmargin  -.4in
	\headheight 12pt \topskip .4in
	\textheight 6.9in \footskip 0pt
	\def\@oddhead{\thepage\hfil\addtocounter{page}{1}\thepage}
	\let\@evenhead\@oddhead	\def\@oddfoot{}	\def\@evenfoot{} }

\def\numberbysection{\@addtoreset{equation}{section}
	\def\theequation{\thesection.\arabic{equation}}}

\def\underline#1{\relax\ifmmode\@@underline#1\else
	$\@@underline{\hbox{#1}}$\relax\fi}

\def\titlepage{\@restonecolfalse\if@twocolumn\@restonecoltrue\onecolum
n
     \else \newpage \fi \thispagestyle{empty}\c@page\z@
	\def\thefootnote{\fnsymbol{footnote}} }

\def\endtitlepage{\if@restonecol\twocolumn \else \newpage \fi
	\def\thefootnote{\arabic{footnote}}
	\setcounter{footnote}{0}}  

\catcode`@=12
\relax

\def\figcap{\section*{Figure Captions\markboth
	{FIGURECAPTIONS}{FIGURECAPTIONS}}\list
	{Figure \arabic{enumi}:\hfill}{\settowidth\labelwidth{Figure
999:}
	\leftmargin\labelwidth
	\advance\leftmargin\labelsep\usecounter{enumi}}}
 \relax
\def\tablecap{\section*{Table Captions\markboth
	{TABLECAPTIONS}{TABLECAPTIONS}}\list
	{Table \arabic{enumi}:\hfill}{\settowidth\labelwidth{Table
999:}
	\leftmargin\labelwidth
	\advance\leftmargin\labelsep\usecounter{enumi}}}
 \relax
\def\reflist{\section*{References\markboth
	{REFLIST}{REFLIST}}\list
	{[\arabic{enumi}]\hfill}{\settowidth\labelwidth{[999]}
	\leftmargin\labelwidth
	\advance\leftmargin\labelsep\usecounter{enumi}}}
 \relax
\makeatletter
\newcounter{pubctr}
\def\publist{\@ifnextchar[{\@publist}{\@@publist}}
\def\@publist[#1]{\list
	{[\arabic{pubctr}]\hfill}{\settowidth\labelwidth{[999]}
	\leftmargin\labelwidth
	\advance\leftmargin\labelsep
	\@nmbrlisttrue\def\@listctr{pubctr}
	\setcounter{pubctr}{#1}\addtocounter{pubctr}{-1}}}
\def\@@publist{\list
	{[\arabic{pubctr}]\hfill}{\settowidth\labelwidth{[999]}
	\leftmargin\labelwidth
	\advance\leftmargin\labelsep
	\@nmbrlisttrue\def\@listctr{pubctr}}}
 \relax
\makeatother
\newskip\humongous \humongous=0pt plus 1000pt minus 1000pt

\newif\ifdtup

\relax

\def\thefootnote{\fnsymbol{footnote}}
\def\be{\begin{equation}}
\def\ee{\end{equation}}
\def\ba{\begin{eqnarray}}
\def\ea{\end{eqnarray}}

\def\S{\Sigma}
\def\s{\sigma}

\def\a{\alpha}

\begin{document}
\renewcommand{\theequation}{\thesection.\arabic{equation}}
\begin{titlepage}
\begin{center}

\hfill CERN-TH.7003/93\\
\hfill hep-th/9309064\\
\vskip .5in

{\large \bf Duality Symmetries and Topology Change in String
Theory}\footnote{
Talk given at the EPS 93 Conference, held at Marseille, July 22-27.
To appear in the Proceedings.}
\vskip .5in

{\bf Elias Kiritsis}
\vskip .1in
{\em Theory Division, CERN\\
 Geneva 23, SWITZERLAND}
\vskip .1in
\end{center}

\vskip .9in

\begin{center} {\bf ABSTRACT } \end{center}
\begin{quotation}\noindent
Duality symmetries for strings moving in non-trivial
spacetime
backgrounds are analysed. It is shown that for backgrounds generated
from
WZW and coset CFT models such duality symmetries are exact to all
orders in
string perturbation theory. Their implications for string dynamics in
non-trivial/singular  spacetimes are discussed.

\end{quotation}
\vskip3.0cm
\vfill
CERN-TH.7003/93 \\
September 1993\\
\end{titlepage}
\vfill
\eject
\def\baselinestretch{1.2}
\baselineskip 16 pt
\noindent

\vskip 4mm
{\bf \noindent Introduction}
\vskip 2.0mm

Strings, being extended objects, sense the target space, into which
they are
embedded, in a different way than point particles.
In a compact space this difference appears because, strings,
except from their local excitations that
mimic point particle behavior (``momentum" modes), have ``winding"
excitations where the string wraps around non-contractible cycles of
the
manifold.
The masses of momentum modes  are inversely proportional to the
volume of the
manifold, whereas those of the winding modes are proportional to the
volume,
since it costs energy in order to stretch the string.
Moreover, the string contains oscillating modes that respond to
background fields differently than the center of mass of the string.
In certain cases, the physics of string propagation remains invariant
under a reorganization of the one string Hilbert space and a specific
change in the background. This symmetry is known as duality.
In the simplest possible example, that of a string moving on a
circle,
it was observed that the spectrum of the theory with radius $R$ and
that with
radius $1/R$ are identical, once we interchange winding and momentum
modes,
[1].

It turns out that such duality symmetries exist (semiclassically) for
all backgrounds  with isometries, [2].
In CFT, some of these symmetries were identified as different abelian
gaugings
of a WZW theory, [3], and this was generalized to abelian
gaugings, [4], and organized into (semi-classical) O(d,d,Z) type
symmetries, [5], mimicking the situation for flat backgrounds.
Moreover, for coset models, such duality symmetries exist also for
backgrounds without any isometries, [3,6].
A careful analysis of the underlying CFT structure, revealed that
most of these semiclassical symmetries (pertaining to compact cosets)
are indeed exact in string theory, and they are intimately related to
the affine Weyl symmetries of the ``parent" theory, the WZW model.

In the non-compact case, the affine Weyl group is not a manifest
symmetry
but it can be shown that a particular kind of duality, axial-vector
duality, [3] is still a symmetry, [7].

At the semiclassical level, provided there is an abelian isometry,
the duality transformation can be effected by gauging this isometry
and adding also
a langrange multiplier coupled to the field strength of the gauge
field, [2].
Integrating out the langrange multiplier, forces the gauge field to
be pure gauge which can, subsequently, be gauged away, giving back
the original model.
On the other hand, one can gauge fix to a unitary gauge and then
integrate out the gauge field (which appears quadratically in the
action).
In this way, a different (dual) sigma model action is obtained (the
measure can be also taken care off, effectively changing the
dilaton).
Modulo global properties, the original and the dual action describe
the same theory.

\vskip 2.0mm
{\bf \noindent Duality in WZW and Coset Models}
\vskip 2.0mm

Applying this procedure to the WZW theory, using the isometries
giving rise
to the Cartan subalgebra of the current algebra, we can verify that
the dual
action is similar to the original one up to periodicities of the
Cartan angles,
 [3].
The dual metric has Taub-Nut type of singularities.
At the level of the Hilbert space, these duality transformations
correspond to
Weyl transformations of the (left$\times$right) current algebra.
Thus, the duality group is $W_{L}\times W_{R} /W_{D}$, where
$W_{L,R}$ are
the left(right) Weyl groups of the (finite) Lie algebra of the WZW
model
and $W_{D}$ is the diagonal Weyl group (whose action corresponds to
reparametrizations of the action) [6].
Duality here is an exact symmetry, since $W_{L}$, $W_{R}$
transformations are symmetries for any highest weight representation
of any compact or non-compact
semi-simple affine Lie algebra, and thus, for the associated WZW
theory, whose spectrum is built from such representations.

{}From the WZW theory we can built other CFTs by projections. The
simplest such projection corresponds to constraining
the affine currents of a subalgebra, known as coset construction,
[8].\footnote{\rm
There are more general projections though, preserving conformal
invariance,
[9].}
The $\s$-model action of coset models is obtained by gauging the
appropriate
subgroup of the WZW model.
Gauging different dual versions of the WZW model, dual versions of
the coset model are obtained. In most cases, this duality exists for
coset actions without isometries, [6].

A special form of duality is obtained for coset models where the
gauged subgroup contains a U(1) factor.
In such a case, one has the option of gauging either the axial or the
vector
subgroup of the original $U(1)_{L}\times U(1)_{R}$ subalgebra.
The $\s$-model actions of these two gauged models are generically
different
but it can be shown that the models are dual to each other, [3].
This type of duality is known as axial-vector duality and at the
semiclassical
level is powerful in generating different types of backgrounds, [4].
It would seem that Weyl symmetry of the affine algebra is enough to
guarantee
that axial-vector duality is an exact symmetry.
The story is more complicated though. It turns out, [6] that the
underlying symmetry of current algebra responsible for axial-vector
duality
is affine-Weyl symmetry, $\hat W_{L}\times \hat W_{R}$.
For integrable (unitary) representations of (compact) affine algebras
the affine Weyl group is a symmetry and so is axial-vector duality.
I conjecture that the the most general (dimension
preserving)\footnote{
\rm By the term ``dimension preserving" I exclude the ``collapse"
phenomenon
that happens for example at WZW of level 1 or rank level duality,
where in both cases the dual theories have a target space with
different dimension.} duality group $D^{g}$ of a compact, {\em
unitary}
$g$-WZW model is isomorphic to $\hat W^{g}_{L}\times \hat W^{g}_{R}
\times \hat A^{g}/W^{g}_{D}$
where $\hat A^{g}$ is the group of external automorphisms of the
current algebra $\hat g$.
The action of $A^{g}$ in $\s$-model language is not yet known.

For compact cosets, the full duality group is obtained by reduction
of the WZW one.
In particular, there are subgroups $D^{h}$ of $D^{g}$ for every
reductive
subalgebra $h\subset g$. The quotient $D^{g}/D^{h}$ is the maximal
duality group of the coset CFT $G/H$.

The full set of duality symmetries of WZW and coset models generates
dualities for other
theories, which are continuously connected to them by marginal
perturbations.
Consider such a theory, which is connected to a WZW (or coset) model
via a marginal perturbation $g\int O_{1,1}$ where $O_{1,1}$ is a
(1,1) operator of the WZW (or coset)  theory.
Self-duality at the WZW (coset) point implies that the line of
theories parametrized by $g$ is equivalent to the one generated by
$\int \tilde O_{1,1}$.
The background interpretation of the two perturbations is generically
different. Thus, in the full moduli space, duality symmetries can be
generated
by the self-duality symmetries at special points (WZW and
cosets).\footnote{
\rm For flat toroidal backgrounds, the full duality symmetry is
generated
this way, [10].}
It is an open question in the non-flat case, if this construction
exhausts all duality symmetries of the moduli space.

The situation for G non-compact (or non-integer level of compact G)
is more complicated and axial-vector equivalence not obvious
(although it still
works semiclassically).
In such a case, the affine Weyl group maps most representations to
different
ones. Thus, a priori, without knowledge of the spectrum of
representations
duality is at stake.
The only way for it to survive is, if the spectrum is organized into
complete
orbits of the affine Weyl group.

It can be shown, [7], that even for non-compact abelian cosets,
axial-vector duality remains an exact symmetry.
I will focus on the simplest non-trivial case, that of
$SL(2,R)/U(1)$,
since it contains all the important ingredients of the general case.

The key step is to construct the $\s$-model description of the line
of
CFTs (parametrized by a positive real number R) generated by the
$\int J^{3}\bar J^{3}$ perturbation of the WZW model.
This $\s$-model can be found by imposing the following three
requirements:

1) It has a $U(1)_{L}\times U(1)_{R}$ chiral symmetry along the hole
line.
\newline\indent
2) At any R the variation of the action is of the form $\int
J^{3}(R)\bar J^{3}(R)$.
\newline\indent
3) At R=1, it reduces to the known $\s$-model action for SL(2,R).

These three requirements fix the $\s$-model action (and measure)
uniquely
and this action is correct to all order in $\alpha$' (in a certain
scheme).
This action can be found in [7].
Around $R=1$, the theories at $1+\delta R$ and $1-\delta R$ are
equivalent
since they are related by a left Weyl transformation in the WZW
theory,
$J^{3}\to -J^{3}$, $\bar J^{3}\to \bar J^{3}$.
This infinitesimal duality transformation propagates along the line
and its
finite form is $R\to 1/R$.

By looking at the form of the action at the end points $(R=0,\infty)$
we observe that at $R=0$ the theory is a direct product of a free
non-compact
boson and the (uncorrelated) vectorial $SL(2,R)/U(1)$ coset model.
At $R=\infty$ the theory also factorizes directly into a free
non-compact boson
and the axial $SL(2,R)/U(1)$ coset model.
In view of the $R\to 1/R$ duality this shows that the axial and
vector
$SL(2,R)/U(1)$ models are equivalent.
This can be generalized to all non-compact abelian cosets, [7].

\vskip 2.0mm
{\bf \noindent Topology change}
\vskip 2.0mm

A slightly modified version of the model discussed in the previous
section
can provide a simple example of smooth topology change in string
theory.
The modification amounts to a blowing up (via a GL(2,R)
transformation)
of the shrinking $S^{1}$.
The line element of the metric is
\be
ds^2(\alpha)=\frac{k}{\Delta(\a)}[(1-\S)d\theta_{1}^2+(1+\S)d\theta_{2
}^2]+
kdx^2
\ee
where
\be
\S=\cos 2x,\;
\Delta=\cos^2 \a (1+\S )+(\cos\a+k\sin\a)^2(1-\S )
\ee
and the SL(2,R) case is obtained for $x\to ix$.
At $\alpha=0$ the metric describes an $S^{3}$ (or pseudo-sphere in
the SL(2,R)
case).
This deforms continuously until $\alpha=\pi/2$ where the metric
becomes
\be
ds^2(\alpha =\pi
/2)=\frac{1}{k}[d\theta_{1}^2+\frac{1+\S}{1-\S}d\theta_{2}^2 ]+kdx^2.
\ee
with the topology of $D_{2}\times S^{1}$, where $D_{2}$ is a 2-disk.
This gives an example of a smooth topology change.
What is also interesting here is that the $\alpha=0$ neighborhood
is mapped via an $O(2,2)$ duality transformation to that around
$\alpha=\pi/2$.
Thus, this maps a region where topology change occurs to one where
topology
is not changed at all.
A similar phenomenon of topology change happens in more complicated
examples of CY compactifications, [11].

\vskip 2.0mm
{\bf \noindent Comments on the Physics of Duality}
\vskip 2.0mm

The main lesson from duality and related symmetries is that the
background
fields do not determine uniquely the spectrum and physics of string
theory.
Duality can be viewed as a tiny (unbroken) part of the huge string
gauge symmetry, whose glory remains obscure to our days.
Another way to state this, is, that different modes of the string
feel
different geometry.
Thus, the background interpretation of string vacuua should be used
with care
in order to ascertain the physics.
The only cases were the description is reliable is the large volume
limit
of compact manifolds, and at the asymptotically flat region of
non-compact manifolds. Once at a region of finite curvature, the
geometrical
description of string theory breaks down.
As we have seen, even topology is not preserved under duality and
there are continuous families
of ground states in string theory where topology changes without the
occurrence of anything catastrophic.

An example of how this type of symmetry can affect string
propagation, can be given (heuristically) as  follows\footnote{\rm
This argument is advanced
in collaboration with C. Kounnas.}.
Consider a string background which is highly curved or even singular
(semiclassically) in a certain region, (the 2-d black hole, [12]
is such an example but one needs to add a few extra dimensions in
order to have non-trivial massive states).
In the asymptotic region, (which is obtained by some
spacetime-depended radius becoming very large), one has quantum
numbers for asymptotic states that
correspond roughly to windings and momenta. Momentum states are the
only low energy states in this region.
Consider a momentum mode travelling towards the high curvature
region.
Its effective mass starts growing as it approaches large curvatures.
At some point it becomes energetically possible for it to decay to
winding
states which, in this region, start having effective masses that are
lower
than momentum modes.
In such backgrounds (unlike flat ones) winding and momentum are not
separately conserved so that such a transition is possible.
The reason for this is that there is a non-trivial dilaton field and
thus, winding and momentum conservation is broken by the screening
operators
which transfer it to discrete states localized at the high curvature
region. An alternative interpretation of this, is that particles
interact
with such localized states loosing momentum (in discrete steps) and
gaining
winding number.

Once such a momentum to winding mode transition happens in the
strongly curved
region, the winding state sees a different geometry, namely the dual
one and thus continues to propagate further into the strong curvature
region
since it feels only the (weak) dual curvature.

This type of picture implies that the physics of string black holes
will be qualitatively different that their classical general
relativity counterparts.
\vfill\eject
\vskip 2.0mm
{\bf \noindent References}
\vskip 2.0mm

\noindent
[1] K. Kikkawa, M Yamazaki, {\it Phys. Lett.} {\bf B149} (1984) 357;
N. Sakai, I. Senda, {\it Prog. Theor. Phys.} {\bf 75} (1986) 692.
\newline\noindent
[2] T. H. Buscher, {\it Phys. Lett.} {\bf B201} (1988) 466.
\newline\noindent
[3] E. Kiritsis, {\em Mod. Phys. Lett.} {\bf A6} (1991) 2871.
\newline\noindent
[4] M. Ro\v cek, E. Verlinde, {\em Nucl. Phys.} {\bf B373} (1992)
630.
\newline\noindent
[5] A. Giveon, M. Ro\v cek, {\em Nucl. Phys.} {\bf B380} (1992) 128.
\newline\noindent
[6] E. Kiritsis, CERN preprint, CERN-TH.6797/93, \newline\noindent
hepth/9302033.
\newline\noindent
[7] A. Giveon, E. Kiritsis, CERN preprint,\newline\noindent
 CERN-TH.6816/93, hepth/9303016.
\newline\noindent
[8] M. Halpern and E. Kiritsis, {\em Mod. Phys. Lett.} {\bf A4}
(1989) 1373.
\newline\noindent
[9] K. Bardakci and M. Halpern, {\em Phys. Rev.} {\bf D3} (1971)
2493;
M. Halpern, {\em Phys. Rev.} {\bf D4} (1971) 2398; P. Goddard, A.
Kent and D. Olive, {\em Phys. Lett.} {\bf B152} (1985) 88.
\newline\noindent
[10] A. Giveon, N. Malkin and E. Rabinovici, {\em Phys. Lett. } {\bf
B238} (1990) 57.
\newline\noindent
[11] P. Aspinwall, B. Greene, D. Morrison, IAS Preprint,
IASSNS-HEP-93-4, hepth/9301043.
\newline\noindent
[12] E. Witten, {\em Phys. Rev.} {\bf D44} (1991) 314.

\end{document}